\documentclass[aps,pra,groupedaddress,showpacs]{revtex4}

\usepackage{graphicx}
\begin{document}

%\tighten

\title{Quantum-state tomography for spin-$l$ systems}

\author{Holger F. Hofmann}
\email{h.hofmann@osa.org}
\author{Shigeki Takeuchi}
\author{}
%\altaffiliation{}
\affiliation{PRESTO, 
Japan Science and Technology Corporation 
(JST)\\
Research Institute for Electronic Science, Hokkaido 
University\\
Kita-12 Nishi-6, Kita-ku, Sapporo 060-0812, Japan}

\date{\today}

\begin{abstract}
We show that the density matrix of a spin-$l$ system can be
described entirely in terms of the measurement statistics
of projective spin measurements along a minimum of 
$4l+1$ different spin directions. It is thus possible to
represent the complete quantum statistics of any $N$-level system 
within the spherically symmetric three dimensional space 
defined by the spin vector. 
An explicit method for reconstructing the density 
matrix of a spin-1 system from the measurement statistics of 
five non-orthogonal spin directions is presented and the 
generalization to spin-$l$ systems is discussed.
\end{abstract}

\pacs{
03.65.Wj, %--State reconstruction, quantum tomography
03.67.-a, %--Quantum information
42.50.-p  %--Quantum optics
}
%%-----(42.50.Dv %--Nonclassical field states) 
%\keywords{}

\maketitle

\section{Introduction}

As rapid progress is being made in the experimental generation
of quantum states, it becomes necessary to develop efficient
methods of characterizing the actual mixed state output of
each new realization. In particular, various types of
optical spin-1 systems have recently been generated using 
parametric downconversion 
\cite{Tse00,Bur02,Lam01,How02,Mai01,Vaz02}. It is therefore 
interesting to consider the measurements necessary to 
properly identify the quantum states of such spin systems.
 
In the most general case, these states can be characterized by 
reconstructing the complete density matrix from 
a sufficiently large set of measurements, a procedure commonly 
referred to as quantum tomography \cite{Whi99,Jam01,The02,Klo01}.
For two-level systems (qubits), quantum tomography is usually
realized by measuring the three orthogonal components of the
Bloch vector represented by the Pauli operators. 
In spin-1/2 systems, the physical meaning of these components
is generally clear. In particular, they represent the
components of the three dimensional Stokes vector in the
commonly studied case of single photon polarization
\cite{Whi99,Jam01}.
In spin-$l$ systems with higher total spin, the connection 
between the much larger number of density matrix elements 
and the physical properties of the system is less clear. 
For abstract N-level systems (qudits), an expansion of the 
density matrix into the generators most closely related to 
the individual density matrix elements has been 
proposed \cite{The02}. However, 
the physical properties corresponding to these operators are
quite different from the spin components observed e.g. in 
Stern-Gerlach or n-photon polarization measurements.

In particular, the recently generated n-photon polarization 
states are usually characterized by photon detection 
measurements in a pair
of orthogonal polarization directions 
\cite{Tse00,Bur02,Lam01,How02}. This corresponds to a 
projective measurement of one component of the three dimensional 
Stokes vector, which is formally equivalent to the three 
dimensional spin vector. The direction of the Stokes vector 
component is determined by the pair of orthogonal polarization 
directions detected in the measurement and can be varied by 
using standard birefringent elements such as quarter wave 
plates and half wave plates. The experimental characterization 
of such optical 
spin-$l$ systems thus corresponds to the measurement of
spin components $\hat{L}_i$ along a set of well defined 
measurement directions $i$. It is therefore desirable
to formulate quantum tomography in terms of the measurement
statistics obtained in this kind of measurements.

In the following, we show how the measurement statistics obtained
in measurements of spin components $\hat{L}_i$ relate to the 
elements of the density matrix. Based on these 
results, a systematic approach to the quantum tomography of 
spin-$l$ systems is developed. We propose a decomposition of the
density matrix into components that reflect the spherical 
symmetry of the spin system and correspond directly to 
well-defined contributions in the experimentally observable
spin statistics. It is shown that measurements along a minimum 
of $4l+1$ spin directions are necessary to 
reconstruct the complete density matrix. 
An explicit method for reconstructing
a spin-1 density matrix from the measurement probabilities
along five non-orthogonal spin directions is derived and 
the extension of this method to general spin-$l$ systems is 
discussed. Since this method can be applied equally well to
small (few level) and large (many level) quantum systems, 
it also provides a useful basis for the discussion of 
decoherence and the transition from quantum to classical
physics.

\section{Measurement statistics of a spin-$l$ system}

Each projective von-Neumann measurement of a spin component
$\hat{L}_i$ results in one of the $2l+1$ eigenvalues $m$
of the quantized spin along the direction corresponding to
$i$. By repeating the measurement a large number of times, it
is possible to determine the probability distribution $p_i(m)$ 
of the measurement outcomes $m$.
The information represented by this probability distribution
can also be expressed in terms of averages of different powers
of $\hat{L}_i$,
\begin{equation}
\label{eq:spinstat}
\langle (\hat{L}_i)^n \rangle = \sum_{m} m^n \; p_i(m).
\end{equation}
The probability distribution over the $2l+1$ possible outcomes
is then uniquely defined by the $2 l$ averages obtained for 
$n=1$ to $n=2 l$, that is, the $2l$ expectation values form
a set of linearly independent parameters describing the 
complete measurement statistics. 

Using this representation of the measurement information
obtained along one spin direction,  
it is now possible to derive the relations between the 
measurement statistics along different spin directions 
by expressing the spin statistics 
of an arbitrary spin direction in terms of the three orthogonal 
spin components $\hat{L}_x$, $\hat{L}_y$, and $\hat{L}_z$.
If the spin direction is given in terms of the horizontal 
and azimuthal angles $\phi$ and $\theta$, the measurement
statistics of $\hat{L}(\phi, \theta)$ then read
\begin{equation}
\label{eq:stats}
\left\langle \left(\hat{L}(\phi, \theta)\right)^n \right\rangle =
\left\langle \left(
\sin(\theta)\cos(\phi) \hat{L}_x
+ \sin(\theta)\sin(\phi) \hat{L}_y
+ \cos(\theta) \hat{L}_z \right)^n \right\rangle.
\end{equation}
Each $n$th order expectation value $\langle (\hat{L}_i)^n \rangle$
can therefore be expanded into expectation values of 
$n$th order products of the three orthogonal spin components.  
Specifically, the linear spin expectation values $(n=1)$ are
defined by the three components of the average spin,
$\langle \hat{L}_x \rangle$, $\langle \hat{L}_y \rangle$,
and $\langle \hat{L}_z \rangle$, while the quadratic spin 
expectation values are given by expectation values such as
$\langle \hat{L}_x \hat{L}_y + \hat{L}_y \hat{L}_x \rangle$,
describing correlations and fluctuations of the spin, and so 
on. The measurement statistics of any measurement direction
can therefore be described by a hierarchy of expectation values
ranging from $n=1$ to $n=2l$.

Although there are an infinite number of measurement directions,
only a finite number of independent parameters is necessary to
describe all $n$th order expectation values. In general,
these independent parameters can be obtained by multiplying
out equation (\ref{eq:stats}) and identifying the contributions
associated with different functions of the angles $\phi$ and 
$\theta$. Alternatively, it is possible to identify these
contributions with elements of the density matrix. As will be
described in more detail below, this corresponds to an expansion
of the density matrix into basis operators $\hat{\lambda}_{n,i}$
that can be expressed as $n$th order polynomial functions of the
spin component operators $\hat{L}_i$. 
Since the density matrix uniquely determines the spin statistics,
the number of independent $n$th order expectation values should 
be equal to the number of independent basis operators 
$\hat{\lambda}_{n,i}$ in the expansion of the density matrix. 
The additional density matrix elements associated with an 
increase of spin from spin-$(l-1/2)$ to spin-$l$ can then be 
identified with the additional $(n=2l)$th order expectation 
values required to describe the spin statistics given by 
equation (\ref{eq:stats}).

The number of parameters defining the density matrix of 
any $N$-level system is $N^2-1$.
A spin-$l$ system is an $N=2l+1$ level system, and the 
number of parameters defining the density matrix is
$4 l (l + 1)$. A spin-$(l-1/2)$ system is an $N=2l$ level 
system, and the number of parameters defining the density 
matrix is $4 l^2 -1$.
The number of additional parameters needed to describe a
spin-$l$ system rather than a spin-$(l-1/2)$ system is 
therefore equal to $4l+1$. Since these additional parameters
characterize the $(n=2l)$th order of the measurement 
statistics given by equation (\ref{eq:stats}), the number 
of independent $n$th order expectation values should be 
equal to $2n+1$.  
Note that this result is indeed consistent with the number of
symmetrically ordered $n$th order products of the
three spin components required to describe 
$\langle \hat{L}_i^n \rangle$, once the constant value of the
total spin length defined by 
$\hat{L}_x^2+\hat{L}_y^2+\hat{L}_z^2=l(l+1)$ has been taken 
into account. Explicit examples for components of the 
$n$th order spin expectation values will be given in sections 
\ref{sec:exp1} and \ref{sec:l>1}. 

\begin{table}
\begin{tabular}{|cllc|ccccc|c|}
\hline
\multicolumn{4}{|c|}{}&\multicolumn{6}{c|}{Components of the}
\\
\multicolumn{4}{|c|}
{System size}&\multicolumn{6}{c|}{
density matrix}
\\[0.1cm]
\multicolumn{4}{|c|}{}& $\langle \hat{L}_i \rangle$ &
$\langle \hat{L}_i^2 \rangle$ & $\langle \hat{L}_i^3 \rangle$ & 
$\langle \hat{L}_i^4 \rangle$ & 
\hspace{0.2cm}$\cdots$\hspace{0.2cm} & total
\\[0.1cm]
\hline &&&&&&&&&
\\[-0.3cm]
& spin-1/2 & (2-level) && 3 & - & - & - &$\cdots$& 3
\\
& spin- 1 & (3-level) && 3 & 5 & - & - &$\cdots$& 8
\\
& spin-3/2 & (4-level) && 3 & 5 & 7 & - &$\cdots$& 15
\\
& spin- 2 & (5-level) && 3 & 5 & 7 & 9 &$\cdots$& 24
\\
& \multicolumn{3}{c|}{$\vdots$} & 
$\vdots$ & $\vdots$ & $\vdots$ & $\vdots$ & & $\vdots$
\\[0.1cm]
\hline
\end{tabular}
\caption{\label{tab:nth}
Reconstruction of the density matrix from n-th order 
expectation values of the spin obtained from projective
measurements of various spin directions $i$. The total 
number of matrix elements necessary for tomography is 
$N^2-1$, where $N=2 l+1$ is the dimensionality of the 
corresponding Hilbert space. 
}
\end{table}

Table \ref{tab:nth} illustrates the distribution of density
matrix parameters. Any spin-$l$ system is characterized
by $2n+1$ expectation values of $n$th order, with $n$ running
from $n=1$ to $n=2l$. That is, the density matrix can always be 
represented by three linear spin averages, five second order 
spin averages, seven third order spin averages, and so on
\cite{foot1}.
Experimentally, each measurement along a given spin direction
determines one $n$th order average for each value of $n$.
Since complete quantum tomography requires the determination 
of $4l+1$ independent contributions to the expectation 
values of order 
$n=2 l$, it is therefore necessary to measure at least 
$4l+1$ different spin directions in order to obtain the 
necessary measurement information \cite{New68}. 
Note that this condition 
is a result of allowing only measurements of spin components 
$\hat{L}_i$. If general von-Neumann measurements were 
possible, $2l+2$ measurements
would be sufficient for quantum tomography. In the case of 
spin component measurements, about half of the information 
obtained is redundant, since it reproduces the results for
lower order expectation values that can be obtained from
fewer spin directions. For example, quantum tomography of
a spin-2 system requires measurements along nine spin directions,
providing nine averages to determine the three linear spin 
expectation values, nine averages to determine the five quadratic
spin expectation values, nine averages to determine the seven
third order spin expectation values and nine averages to 
determine the nine fourth order expectation values. 
Thus it is only the need to determine the complete $(n=2l)$th 
order spin statistics that makes it necessary to measure a 
total of $4l+1$ spin directions $\hat{L}_i$. 

Note that the precise choice of measurement directions is
not very critical, since the only requirement for obtaining
the complete spin statistics is that $2n+1$ of the $4l+1$
averages obtained from the $n$th order statistics are linearly 
independent. In some special cases, the information obtained
from $4l+1$ different measurement directions may not be
sufficient to reconstruct the complete density matrix 
because the choice of measurement directions 
coincides with a symmetry in the $(n=2l)$th order expectation
values. For example, the 2nd order information obtained for 
$\hat{L}_z$ is already obtained from measurements of 
$\hat{L}_x$ and $\hat{L}_y$, since 
$\hat{L}_x^2+\hat{L}_y^2+\hat{L}_z^2=l(l+1)$.  
However, even a small tilt of one of the measurement
axes can fix this problem by providing additional information.
Therefore, almost any choice of $4l+1$ measurement directions 
will be sufficient for quantum tomography. To minimize errors, 
it may be useful to keep the angles between different directions 
as large as possible \cite{Klo01}. 
However, the choice of $4l+1$ measurement 
directions used for a complete reconstruction of the
density matrix can generally be quite arbitrary, and each 
specific tomography protocol merely represents one example 
out of an infinity of equally valid possibilities.

In order to obtain an explicit description of the density matrix
in terms of $n$th order spin statistics, it is necessary to
identify the contributions of different order $n$ in the 
density matrix. This can be achieved by expanding the density 
matrix using an appropriate operator basis. 
In general, there are infinitely many expansions of the
density matrix into orthonormal basis operators. For reasons
of mathematical simplicity, the most common choice is that 
given by the generators of the SU($N$) algebra 
\cite{Hio81,Mah98}. However, these generators do not represent 
the spherical symmetry of the spin-$l$ system. We therefore
propose an alternative expansion of the spin-$l$ density matrix 
$\hat{\rho}_l$ that is based on the different orders $n$ of the
spin statistics, given by basis operators $\hat{\lambda}_{n,i}$,
where $n$ represents the lowest order of the spin statistics to
which $\hat{\lambda}_{n,i}$ contributes, and $i$ is the index 
of the component within this order, running from $1$ to $2n+1$
for each value of $n$. The conditions for orthogonality and
for the normalization of this basis then read
\begin{eqnarray}
\label{eq:generate}
\mbox{Tr}\{\hat{\lambda}_{n,i} \} &=& 0
\nonumber \\
\mbox{Tr}\{\hat{\lambda}_{n,i}\hat{\lambda}_{m,j} \} 
&=& 2 \; \delta_{n,m}\,\delta_{i,j}.
\end{eqnarray}
Using this complete operator basis, it is possible to expand
the spin-$l$ density matrix $\hat{\rho}_l$ in terms of 
expectation values of the spin statistics as
\begin{equation}
\label{eq:expand}
\hat{\rho}_l = \frac{1}{2l+1} \hat{1} + \frac{1}{2} 
\sum_{n=1}^{2l} \left(\sum_{i=1}^{2n+1} 
\langle \hat{\lambda}_{n,i} \rangle \; \hat{\lambda}_{n,i}
\right).
\end{equation}
It is thus possible to formulate the density matrix entirely
in terms of the measurement statistics of the three dimensional
spin vector, reflecting the analysis of the measurement
statistics described by equation (\ref{eq:stats}) given in 
table \ref{tab:nth}. Specifically, 
the $(N^2-1)$-dimensional basis is divided into groups of 
$n$th order products of the spin operators with $2n+1$ 
operators each, describing the separate $n$th order 
contributions to the spin statistics.
In the following, we show how such a basis of 
three linear and five quadratic spin operators can be 
defined for the specific case of $l=1$.

\section{Expansion of the spin-1 density matrix}
\label{sec:exp1}

The case of $l=1$ provides the most simple example of 
non-linear contributions to the spin statistics.
Moreover, optical spin-1 systems have already been
realized by two photon polarization states \cite{Tse00,Bur02,
Lam01,How02} 
or by the orbital angular momentum of single photons 
\cite{Mai01,Vaz02}. The following procedure for quantum tomography
of a spin-1 system may therefore be particularly
useful in the characterization of such experimental 
results.

For $l=1$, the linear spin components already fulfill
the conditions defined by equations (\ref{eq:generate}).
The remaining five basis operators can then be constructed
using 2nd-order operator products. One such set of quadratic
operators that fulfills the relations defined by 
(\ref{eq:generate}) is \cite{Hof03}
\begin{eqnarray}
\label{eq:definitions}
\hat{Q}_{ij} &=& \hat{L}_i\hat{L}_j+\hat{L}_j\hat{L}_i
\nonumber \\
\hat{S}_{xy} &=& \hat{L}_x^2 - \hat{L}_y^2
\nonumber \\
\hat{G}_{z} &=& - \frac{1}{\sqrt{3}} 
\left(\hat{L}_x^2 + \hat{L}_y^2 - 2 \hat{L}_z^2 \right).
\end{eqnarray}
In this basis, the generalized eight dimensional Bloch vector
can be separated into a three dimensional linear part and
a five dimensional quadratic part, given by
\begin{equation}
\label{eq:lambda}
\begin{array}{lcrclcr}
\hat{\lambda}_{1,1} &=& \hat{L}_x &&
\hat{\lambda}_{2,1} &=& \hat{S}_{xy}
\\
\hat{\lambda}_{1,2} &=& \hat{L}_y &&
\hat{\lambda}_{2,2} &=& \hat{Q}_{xy}
\\
\hat{\lambda}_{1,3} &=& \hat{L}_z &&
\hat{\lambda}_{2,3} &=& \hat{Q}_{yz}
\\
 & & &\hspace{1.5cm}&\hat{\lambda}_{2,4} &=& \hat{Q}_{zx}
\\
 & & &            &\hat{\lambda}_{2,5} &=& \hat{G}_{z}.
\end{array}
\end{equation}
In the $\hat{L}_z$-basis, the matrix elements of these
operators read
\begin{eqnarray}
\label{eq:matrix}
\hat{\lambda}_{1,1} = \frac{1}{\sqrt{2}}
\left[ 
\begin{array}{ccc}
0&1&0 \\ 1&0&1 \\ 0&1&0
\end{array}
\right] &\hspace{1.5cm}& 
\hat{\lambda}_{2,1} = \hspace{0.6cm}
\left[ 
\begin{array}{ccc}
0&0&1 \\ 0&0&0 \\ 1&0&0
\end{array}
\right]
\nonumber \\
\hat{\lambda}_{1,2} = \frac{1}{\sqrt{2}}
\left[ 
\begin{array}{ccc}
0&-i&0 \\ i&0&-i \\ 0&i&0
\end{array}
\right] && 
\hat{\lambda}_{2,2} = \hspace{0.6cm}
\left[ 
\begin{array}{ccc}
0&0&-i \\ 0&0&0 \\ i&0&0
\end{array}
\right]
\nonumber \\
\hat{\lambda}_{1,3} = \hspace{0.8cm}
\left[ 
\begin{array}{ccc}
1&0&0 \\ 0&0&0 \\ 0&0&-1
\end{array}
\right] && 
\hat{\lambda}_{2,3} = \frac{1}{\sqrt{2}}
\left[ 
\begin{array}{ccc}
0&-i&0 \\ i&0&i \\ 0&-i&0
\end{array}
\right]
\nonumber \\ && 
\hat{\lambda}_{2,4} = \frac{1}{\sqrt{2}}
\left[ 
\begin{array}{ccc}
0&1&0 \\ 1&0&-1 \\ 0&-1&0
\end{array}
\right]
\nonumber \\ &&
\hat{\lambda}_{2,5} = \frac{1}{\sqrt{3}}
\left[ 
\begin{array}{ccc}
1&0&0 \\ 0&-2&0 \\ 0&0&1
\end{array}
\right].
\end{eqnarray}
The expectation values of these eight operators characterize 
the density matrix in terms of the linear and the quadratic
measurement statistics given by equation (\ref{eq:stats}). 
The linear expectation values along any spin direction
are given by the three expectation values of 
$\hat{\lambda}_{1,i}$,
\begin{equation}
\langle \hat{L}(\phi, \theta) \rangle
= \sin(\theta)\cos(\phi) \langle \hat{\lambda}_{1,1} \rangle
+ \sin(\theta)\sin(\phi) \langle \hat{\lambda}_{1,2} \rangle
+ \cos(\theta) \langle \hat{\lambda}_{1,3} \rangle,
\end{equation}
and the quadratic expectation values are given by the 
five expectation values of $\hat{\lambda}_{2,i}$,
\begin{eqnarray}
\langle \hat{L}(\phi, \theta)^2 \rangle
= \frac{2}{3} &+& \frac{1}{2} \sin^2(\theta) \cos(2 \phi)
\langle \hat{\lambda}_{2,1} \rangle
+ \sin^2(\theta) \sin(\phi) \cos(\phi) 
\langle \hat{\lambda}_{2,2} \rangle
+ \sin(\theta)\cos(\theta) \sin(\phi) 
\langle \hat{\lambda}_{2,3} \rangle
\nonumber \\ &+& \sin(\theta)\cos(\theta) \cos(\phi) 
\langle \hat{\lambda}_{2,4} \rangle
+ \frac{1}{\sqrt{3}} (1-\frac{3}{2} \sin^2(\theta))
\langle \hat{\lambda}_{2,5} \rangle.
\end{eqnarray} 
Using this relation, it is possible to determine the 
correct expectation values of all five 2nd order basis
operators $\hat{\lambda}_{2,i}$ 
from the quadratic expectation values 
of five independent measurement directions. 
Together with the linear expectation values, the results 
of these five measurements then define the complete density 
matrix,
\begin{equation}
\hat{\rho}_{l=1} =
\frac{1}{3} \hat{1} \;+\; \underbrace{\frac{1}{2}\sum_{i=1}^3
\langle \hat{\lambda}_{1,i} \rangle \hat{\lambda}_{1,i}
}_{\mbox{from}\hspace{0.3cm}\langle \hat{L}(\phi, \theta) \rangle}
\;+\; \underbrace{\frac{1}{2}\sum_{i=1}^5
\langle \hat{\lambda}_{2,i} \rangle \hat{\lambda}_{2,i}
}_{\mbox{from}\hspace{0.3cm}\langle \hat{L}(\phi, \theta)^2 \rangle}
.
\end{equation}
An explicit procedure for quantum tomography can now be formulated
by choosing a set of five measurement directions. The expectation
values of the basis operators $\hat{\lambda}_{n,i}$ can then
be expressed in terms of the measurement probabilities along
the five measurement directions.

\section{Quantum tomography of the spin-1 system based
on the measurement statistics of five spin directions}

For $l=1$, each measurement along a given spin direction
$\hat{L}_i$ has three possible outcomes, 
$m=\pm 1$ and $m=0$. In the 
experimentally relevant case of two-photon polarization
\cite{Tse00,Bur02,Lam01,How02},
these measurement outcomes correspond to the detection
of two horizontally polarized photons $(m=+1)$, 
two vertically polarized photons $(m=+1)$, 
and one photon each in horizontal and in vertical
polarization $(m=0)$, where the component of the Stokes 
vector is selected by appropriate rotations of the polarization
using standard linear optics elements. The measurement statistics
given by the probabilities $p_i(m)$ can then be identified 
with normalized coincident count rates at the detectors.

As discussed above, five measurement settings are necessary 
to perform quantum tomography. 
A particularly simple choice of the five measurement directions 
for the spin-1 system is given by 
\begin{equation}
\label{eq:directions}
\begin{array}{lll}
\hat{L}_1 = \hat{L}_x, & \hat{L}_2 = \hat{L}_y, &
\hat{L}_3 = \frac{1}{\sqrt{2}}\left(\hat{L}_x + \hat{L}_y\right),
\\
\hat{L}_4 = \frac{1}{\sqrt{2}}\left(\hat{L}_y + \hat{L}_z\right),
&
\hat{L}_5 = \frac{1}{\sqrt{2}}\left(\hat{L}_z + \hat{L}_x\right).
&
\end{array}
\end{equation}
In the case of two-photon polarization, the first three spin 
directions can be identified with linear polarization rotated by
zero for $\hat{L}_1$, by $\pi/4$ for $\hat{L}_2$,
and by $\pi/8$ for $\hat{L}_3$. The remaining two directions
then represent elliptical polarizations, with the main axes
along $\pm \pi/4$ for $\hat{L}_4$, and the main axes along angles
of zero and $\pi/2$ for $\hat{L}_5$. Using this set of measurement
settings, it is now possible to explicitly identify the 
expectation values $\langle \hat{\lambda}_{n,i}\rangle$
that define the density matrix with the corresponding 
measurement probabilities $p_i(m)$.

In general, the measurement probabilities of a spin-1 system
along a given direction $i$ are related to the $n$th order 
expectation values of the corresponding spin component 
$\hat{L}_i$ by
\begin{eqnarray}
\label{eq:spin1}
\langle \hat{L}_i \rangle &=& p_i(+1) - p_i(-1)
\nonumber
\\
\langle \hat{L}_i^2 \rangle &=& 1 - p_i(0).
\end{eqnarray}
The quadratic terms of the spin statistics are thus entirely 
determined by the measurement probabilities $p_i(0)$ for 
a measurement value of zero spin along the measurement
direction $i$. The five expectation values $\hat{\lambda}_{2,i}$ 
defining the quadratic components of the density matrix can 
therefore be obtained from the five measurement probabilities 
$p_i(0)$ along the spin directions $i=1$ to $i=5$ defined by 
(\ref{eq:directions}). The relations between the measurement
statistics and the expectation values of the corresponding
basis operators then reads
\begin{eqnarray}
\langle \hat{\lambda}_{2,1} \rangle = \langle \hat{S}_{xy} \rangle 
&=& - \left(p_1(0) - p_2(0)\right)
\nonumber \\
 \langle \hat{\lambda}_{2,2} \rangle = \langle \hat{Q}_{xy} \rangle 
&=& p_1(0) + p_2(0) - 2 p_3(0)
\nonumber \\
 \langle \hat{\lambda}_{2,3} \rangle = \langle \hat{Q}_{yz} \rangle 
&=& p_1(0) - 2 p_4(0) + 1
\nonumber \\
 \langle \hat{\lambda}_{2,4} \rangle = \langle \hat{Q}_{zx} \rangle 
&=& p_2(0) - 2 p_5(0) + 1
\nonumber \\
 \langle \hat{\lambda}_{2,5} \rangle = \langle \hat{G}_{z} \rangle 
&=& \sqrt{3}\left(p_1(0) + p_2(0) - \frac{2}{3}\right).
\end{eqnarray}
The measurement probabilities for $m=0$ thus determine
five of the eight coefficients in the expansion of the density
matrix given by equation (\ref{eq:expand}). The remaining 
three coefficients can be obtained from the linear 
expectation values given by the differences between 
$p_i(+1)$ and $p_i(-1)$, e.g.
\begin{eqnarray}
\langle \hat{\lambda}_{1,1} \rangle = \langle \hat{L}_{x} \rangle 
&=& p_1(+1) - p_1(-1)
\nonumber \\
 \langle \hat{\lambda}_{1,2} \rangle = \langle \hat{L}_{y} \rangle 
&=& p_2(+1) - p_2(-1)
\nonumber \\
 \langle \hat{\lambda}_{1,3} \rangle = \langle \hat{L}_{z} \rangle 
&=& -\sqrt{2}\left((p_3(+1) - p_3(-1)) - 
(p_4(+1) - p_4(-1)) - (p_5(+1) - p_5(-1))
\right).
\end{eqnarray}
Since the results of five measurement directions are used to
determine only three parameters, there are two relations
between the measurement results that should be approximately
fulfilled if the measurement error is low. These relations
can be written as
\begin{eqnarray}
\label{eq:predict}
(p_1(+1) - p_1(-1)) &\approx&
\frac{1}{\sqrt{2}}\left((p_3(+1) - p_3(-1))-(p_4(+1) - p_4(-1))
+(p_5(+1) - p_5(-1))\right)
\nonumber \\
(p_2(+1) - p_2(-1)) &\approx&
\frac{1}{\sqrt{2}}\left((p_3(+1) - p_3(-1))+(p_4(+1) - p_4(-1))
-(p_5(+1) - p_5(-1))\right).
\end{eqnarray}
Effectively, the measurement results for the spin directions
three to five can be used to predict the results for spin
directions one and two. Relations (\ref{eq:predict})
thus illustrate the application of quantum tomography
to the prediction of further measurements on the same system.
The experimental differences between this prediction and the 
actual outcome of the measurements may therefore provide a
realistic estimate of the errors limiting the reliability of
quantum tomography in practical applications.

Since the eight parameters $\langle \hat{\lambda_{n,i}} \rangle$ 
completely define the density matrix, they can be used as
an alternative representation of the quantum state, just like
the three dimensional Bloch vector for two level systems.
Any other representation of the density matrix can then 
be obtained from equation (\ref{eq:expand}), if the 
representations of each element of the basis is known.
For example, the density matrix elements in the 
$\hat{L}_z$-basis given by equations (\ref{eq:matrix})
can be used to express any density matrix defined by the
expectation values $\langle \hat{\lambda_{n,i}} \rangle$ as
{\small
\begin{eqnarray}
\label{eq:lz}
\lefteqn{
\hat{\rho}_{l=1}=
}
\nonumber \\[0.2cm] && \hspace{-0.3cm}
\left [
\begin{array}{ccc}
\frac{1}{3}+\frac{1}{2} \langle \hat{\lambda}_{1,3} \rangle
+ \frac{1}{2 \sqrt{3}}\langle \hat{\lambda}_{2,5} \rangle
&
\frac{1}{2 \sqrt{2}}\left( \langle \hat{\lambda}_{1,1} \rangle
- i \langle \hat{\lambda}_{1,2} \rangle
- i \langle \hat{\lambda}_{2,3} \rangle
+ \langle \hat{\lambda}_{2,4} \rangle
\right)
&
\frac{1}{2}\left( \langle \hat{\lambda}_{2,1} \rangle
- i \langle \hat{\lambda}_{2,2} \rangle \right)
\\
\frac{1}{2 \sqrt{2}}\left( \langle \hat{\lambda}_{1,1} \rangle
+ i \langle \hat{\lambda}_{1,2} \rangle
+ i \langle \hat{\lambda}_{2,3} \rangle
+ \langle \hat{\lambda}_{2,4} \rangle
\right)
&
\frac{1}{3} - \frac{1}{\sqrt{3}} \langle \hat{\lambda}_{2,5} \rangle
&
\frac{1}{2 \sqrt{2}}\left( \langle \hat{\lambda}_{1,1} \rangle
- i \langle \hat{\lambda}_{1,2} \rangle
+ i \langle \hat{\lambda}_{2,3} \rangle
- \langle \hat{\lambda}_{2,4} \rangle
\right)
\\
\frac{1}{2}\left( \langle \hat{\lambda}_{2,1} \rangle
+ i \langle \hat{\lambda}_{2,2} \rangle \right)
&
\frac{1}{2 \sqrt{2}}\left( \langle \hat{\lambda}_{1,1} \rangle
+ i \langle \hat{\lambda}_{1,2} \rangle
- i \langle \hat{\lambda}_{2,3} \rangle
- \langle \hat{\lambda}_{2,4} \rangle
\right)
&
\frac{1}{3} - \frac{1}{2} \langle \hat{\lambda}_{1,3} \rangle
+ \frac{1}{2 \sqrt{3}}\langle \hat{\lambda}_{2,5} \rangle
\end{array}
\right ].
\nonumber \\[0.2cm]
\end{eqnarray}
}
The quantum coherences between different eigenstates of
$\hat{L}_z$ can thus be expressed in terms of averages 
involving the other two spin components, $\hat{L}_x$ and
$\hat{L}_y$. In fact, this relationship between quantum
coherence in the $\hat{L}_z$-basis and the orthogonal
spin components $\hat{L}_x$ and $\hat{L}_y$ can be used to
systematically construct density matrix decompositions for 
higher spins, as we will show in the following.

\section{Quantum statistics for $l>1$}
\label{sec:l>1}

%---
In order to generalize the construction of a convenient
set of basis operators $\hat{\lambda}_{n,i}$ to arbitrarily
large spins, it is useful to organize the basis operators
according to their density matrix elements in the 
$\hat{L}_z$-basis. 
Such an organization is already indicated by the example
for spin-1 given in equation (\ref{eq:lz}). In this example, 
each group of density matrix elements 
$\mid m \rangle\langle m^\prime \mid$ with the same order 
of coherence $|m^\prime-m|$ in $\hat{L}_z$ depends on a well 
defined subset of the operators $\hat{\lambda}_{n,i}$. 
This organization of basis operators can be generalized 
to arbitrary spins $l$ by formulating the operator basis 
$\hat{\lambda}_{n,i}$ for each spin value in such a way that 
the matrix elements of each operator are non-zero for only one 
value of $|m^\prime-m|$. The relationship between 
the hierarchy of spin expectation values 
$\langle \hat{\lambda}_{n,i}\rangle$ and the density matrix 
elements in the $\hat{L}_z$-basis is then as shown in table
\ref{tab:lzmat}. Specifically, the $2l+1$ diagonal matrix 
elements ($|m^\prime-m|=0$) are determined by the set of 
$2l+1$ operator expectation values 
$\langle \hat{\lambda}_{n,(2n+1)} \rangle$ that can be obtained 
from the measurement statistics of $\hat{L}_z$ according to 
equation (\ref{eq:spinstat}). Likewise, 
the $4l$ off-diagonal elements with $|m^\prime-m|=1$
can be determined by the set of $4l$ operator expectation
values, $\langle \hat{\lambda}_{n,2n} \rangle$ and 
$\langle \hat{\lambda}_{n,2n-1} \rangle$, and so on.
Note that the $4l-2$ off-diagonal elements with $|m^\prime-m|=2$ 
do not include any first order spin statistics, since the linear
spin components only have matrix elements up to $|m^\prime-m|=1$.
\begin{table}
\begin{tabular}{|c|ccccccc|c|}
\hline
& \multicolumn{7}{|c|}{Density matrix elements} &
\\[0.2cm]
spin & 
$|m^\prime-m|$ & $|m^\prime-m|$ &
$|m^\prime-m|$ & $|m^\prime-m|$ & 
$|m^\prime-m|$ & $|m^\prime-m|$ & 
\hspace{0.2cm}$\ldots$ \hspace{0.2cm}
&  
\\[-0.1cm]
statistics & 
$=0$ & $=1$ &
$=2$ & $=3$ & 
$=4$ & $=5$ & 
& total 
\\[0.2cm]
\hline
&&&&&&&&
\\[-0.2cm]
$\langle \hat{\lambda}_{1,i} \rangle$ &
1 & 2 & - & - & - & - & $\cdots$ & 3 
\\
$\langle \hat{\lambda}_{2,i} \rangle$ &
1 & 2 & 2 & - & - & - & $\cdots$ & 5 
\\
$\langle \hat{\lambda}_{3,i} \rangle$ &
1 & 2 & 2 & 2 & - & - & $\cdots$ & 7 
\\
$\langle \hat{\lambda}_{4,i} \rangle$ &
1 & 2 & 2 & 2 & 2 & - & $\cdots$ & 9
\\
$\langle \hat{\lambda}_{5,i} \rangle$ &
1 & 2 & 2 & 2 & 2 & 2 & $\cdots$ & 11
\\
$\vdots$ & 
$\vdots$ & $\vdots$ & $\vdots$ & $\vdots$ & $\vdots$ & $\vdots$
& & $\vdots$
\\
total & 
$2l+1$ & $4l$ & $4l-2$ & $4l-4$ & $4l-6$ & $4l-8$ 
& $\cdots$ 
& $4l(l+1)$
\\[0.2cm]
\hline
\end{tabular}
\caption{\label{tab:lzmat}
Relation between the density matrix elements 
$\langle m^\prime \mid \hat{\rho} \mid m \rangle$ in the 
$\hat{L}_z$-basis and the corresponding $n$th-order basis
operators representing the spin statistics.
}
\end{table}
Likewise, the $4l-4$ off-diagonal elements with $|m^\prime-m|=3$ 
do not include any linear or quadratic spin statistics. 
In general, an expectation value of at least $|m^\prime-m|$th 
order is necessary to describe the effect of a
corresponding off-diagonal element in the spin statistics. 

A convenient way to construct the two operators 
$\hat{\lambda}_{n,1}$ and $\hat{\lambda}_{n,2}$ with 
non-zero density matrix elements of maximal coherence
$|m-m^\prime|=n$ is to apply the non-hermitian spin 
operator $\hat{L}_x+i \hat{L}_y$.
This operator only has non-zero matrix elements with 
$m^\prime-m=+1$. Specifically,
\begin{equation}
\left(\hat{L}_x+i \hat{L}_y\right) \mid m \rangle
= \sqrt{(l-m)(l+m+1)} \mid m+1 \rangle.
\end{equation}
Consequently, $(\hat{L}_x+i \hat{L}_y)^n$ has only matrix
elements with $m^\prime-m=n$. It is therefore possible to
generate the $n$th order basis operators $\hat{\lambda}_{n,1}$
and $\hat{\lambda}_{n,2}$ with matrix elements of 
$|m^\prime-m|=n$ from the normalized hermitian components
of $(\hat{L}_x+i \hat{L}_y)^n$,
\begin{eqnarray}
\label{eq:lpower}
\hat{\lambda}_{n,1} &=& \frac{(\hat{L}_x+i \hat{L}_y)^n
+(\hat{L}_x-i \hat{L}_y)^n}
{\sqrt{\mbox{Tr}\{(\hat{L}_x+i \hat{L}_y)^n
(\hat{L}_x-i \hat{L}_y)^n\}}}
\nonumber
\\
\hat{\lambda}_{n,2} &=& \frac{-i\left((\hat{L}_x+i \hat{L}_y)^n
-(\hat{L}_x-i \hat{L}_y)^n\right)}
{\sqrt{\mbox{Tr}\{(\hat{L}_x+i \hat{L}_y)^n
(\hat{L}_x-i \hat{L}_y)^n\}}}.
\end{eqnarray}
Starting from these definitions of basis operators, the complete 
set of basis operators may be constructed, e.g. by multiplying
the operators $\hat{\lambda}_{n,1/2}$ with different 
powers of $\hat{L}_z$ and/or $\hat{L}_x^2+\hat{L}_y^2$ to obtain
higher order contributions with the same coherence $|m^\prime-m|$ 
in the density matrix. The precise factors can be determined
using the requirements for orthogonality and normalization
given by equation (\ref{eq:generate}).
It is then possible to construct a complete orthonormal 
operator basis for any spin-$l$ system.

By establishing the relation between spin statistics and
coherence in the density matrix, equation (\ref{eq:lpower}) 
also illustrates the physical meaning of quantum coherence 
in $\hat{L}_z$. In particular, it is worth noting that the 
greater the difference $|m^\prime-m|$ between the 
$\hat{L}_z$-eigenvalues
of the states that are in a coherent superposition, the
higher the order of the spin expectation values in 
$\hat{L}_x$ and $\hat{L}_y$ that is needed to identify 
this coherence in the measurement statistics. 
The spin correlation hierarchy 
presented in tabel \ref{tab:lzmat} may thus provide a key to
understanding the non-classical effects associated with 
quantum superpositions in arbitrarily large physical systems.

\section{Non-classical correlations and decoherence}

The highest possible value for $|m^\prime-m|$ in the density 
matrix is obtained for quantum coherence between the extremal
$\hat{L}_z$-eigenstates $\mid m=+l \rangle$ and 
$\mid m=-l \rangle$. According to tabel \ref{tab:lzmat},
the two matrix elements describing this coherence correspond
to the two $(2l)$th order spin expectation values.
These expectation values can be constructed explicitly
using equation (\ref{eq:lpower}) and include only products
of $\hat{L}_x$ and $\hat{L}_y$. 
In order to characterize a coherent superposition of the
$\hat{L}_z$-eigenstates $\mid m=+l \rangle$ and 
$\mid m=-l \rangle$, it is 
therefore necessary to evaluate the $(2l)$th order spin
statistics in the xy-plane. For all orders lower than $2l$, 
the spin statistics obtained in the measurements of such a 
coherent superposition are identical to those of an 
incoherent mixture of $\mid m=+l \rangle$ and 
$\mid m=-l \rangle$. 

This observation has significant 
implications for the identification of strong non-classical
effects in large quantum systems. 
At sufficiently high values of $l$, the superposition 
of $\mid m=+l \rangle$ and $\mid m=-l \rangle$ is
a cat-state like superposition of two macroscopically
distinguishable states. It is therefore interesting 
to know that the effects of this superposition appear
only in the highest order expectation value of the
spin statistics. While the lower order expectation values
are very easy to measure since only very few measurement
directions are required and the measurement errors
tend to average out, the highest order expectation values
can only be determined from sufficiently precise measurement
results of at least $4l+1$ measurement directions.
Effectively, the highest order expectation values represent
a measurement resolution at the quantum level, providing the
information necessary to resolve the precise eigenvalues
of the spin components \cite{foot2}.
This means that the $(2l)$th order spin statistics is very
sensitive to errors of $\pm 1$ in the spin measurements.
In other words, the smallest measurement errors are sufficient
to make the effects of the cat-state like superposition between 
$\mid m=+l \rangle$ and $\mid m=-l \rangle$ disappear.
We can therefore conclude that the actual non-classical 
properties of a superposition of macroscopically 
distinguishable states can only be observed in the microscopic
details of the measurement statistics. 
It is therefore not surprising that decoherence quickly 
wipes out such tiny details. 

For a more precise evaluation of decoherence and measurement
precision, it is useful to consider the case of isotropic 
decoherence caused by spin diffusion due to random rotations.
The time evolution of the density matrix caused by this kind of 
decoherence can be described by 
\begin{equation}
\frac{d}{dt} \hat{\rho}_l = 
- \Gamma \sum_{i=x,y,z} \left(
\frac{1}{2} \hat{L}_i^2\hat{\rho} + 
\frac{1}{2} \hat{\rho}\hat{L}_i^2 - 
\hat{L}_i\hat{\rho}\hat{L}_i
\right).
\end{equation}
Using the well known commutation relations of the spin
operators, it is possible to calculate the relaxation dynamics
of the $n$th-order expectation values of the spin. 
For the non-Hermitian operators $(\hat{L}_x+i \hat{L}_y)^n$
the result reads
\begin{equation}
\frac{d}{dt} \langle (\hat{L}_x+i \hat{L}_y)^n \rangle =
-\frac{n(n+1)}{2} \Gamma 
\langle (\hat{L}_x+i \hat{L}_y)^n \rangle.
\end{equation}
Since the relaxation of the spin is isotropic, all $n$th
order contributions to the expansion of the density matrix
should relax at the same rate. The effect of isotropic 
decoherence therefore reduces each $n$th order parameter
$\langle \hat{\lambda}_{n,i} \rangle$ of the density matrix 
expansion (\ref{eq:expand}) by a decoherence factor of
$\exp[-\Gamma t n(n+1)/2]$, and the time evolution of the
density matrix can be written as
\begin{equation}
\label{eq:deco}
\hat{\rho}(t) = \frac{1}{2l+1} \hat{1}
+\sum_{n=1}^{2l}
\exp\left[-\Gamma t \frac{n(n+1)}{2}\right] \; 
\left(
\sum_{i=1}^{2n+1} 
\langle \hat{\lambda}_{n,i} \rangle_{t=0}\; \hat{\lambda}_{n,i}
\right).
\end{equation}
The expansion of the density matrix $\hat{\rho}_l$ into an 
operator basis $\hat{\lambda}_{n,i}$ based on the different
orders of the spin statistics therefore greatly simplifies
the description of any isotropic errors in the preparation 
and manipulation of spin states.

Since the decoherence effects described by equation 
(\ref{eq:deco}) arise from spin diffusion, it is also 
possible to identify $\Gamma t$ with an increasing
uncertainty in the spin direction, 
$\Gamma t = \delta\theta\,^2/2$. The result of 
equation (\ref{eq:deco}) can then be used to
estimate the errors caused by a misalignment of 
the measurement direction.
Specifically, an alignment error of $\delta\theta$ will
reduce the expectation values observed for
the $n$th order spin statistics by
a factor of $\exp[- \delta\theta\;^2 n(n+1)/4]$.
To obtain at least $\exp[-0.25]=78\%$ of the original 
expectation value at orders $n \gg 1$ of the spin 
statistics, the errors of the spin alignment have to be smaller
than $\delta \theta = 1/n$.
The precision in the alignment of the spin direction 
necessary to obtain the $n$th order statistics 
is thus proportional to $1/n$ and the requirement for 
observing evidence of a cat-like superpositions in 
spin-$l$ systems is an angular resolution of 
$\delta \theta < 1/(2l)$.

\section{Entanglement statistics and general spin networks}

The formalism developed above can also be applied 
to entangled spin-$l$ systems. In this case, 
the density matrix of the total system is obtained
by evaluating the correlations between measurements 
of the local spin components. 
Specifically, the joint quantum state of a spin-$l_A$ 
system A and a spin-$l_B$ system B can be determined
by simultaneously measuring spin components 
$\hat{L}_i(A)$ in A and spin components 
$\hat{L}_i(B)$ in B, obtaining the joint probabilities 
$p_{ij}(m_A,m_B)$ of each measurement outcome. 
The correlated spin statistics can then be expressed 
in terms of the expectation values
\begin{equation}
\label{eq:2stat}
\langle (\hat{L}_i(A))^{n_A}\otimes(\hat{L}_i(A))^{n_B}\rangle
= \sum_{m_A,m_B} m_A^{n_A}\;m_B^{n_B}\;p_{ij}(m_A,m_B).
\end{equation}
It is then possible to analyze the spin statistics according
to the local order $n_A$ and $n_B$, where the total number
of independent components required to characterize each order 
is given by the product $(2n_A+1)(2n_B+1)$.
Note that in this case, $n_A=0$ and $n_B=0$ have to be included
in order to describe the local spin statistics of each system.
Consequently, the lowest order expectation values are
given by $(n_A=1,n_B=0)$ and $(n_A=0,n_B=1)$, with three 
independent components each. The second order expectation 
values ($n_A+n_B=2$) are given by five components for 
$(n_A=2,n_B=0)$, nine components for $(n_A=1,n_B=1)$, 
and five components for $(n_A=0,n_B=2)$. The highest order
contribution to the correlated spin statistics is then
given by $(n_A=2l_A,n_B=2l_B)$, with a total of 
$(4l_A+1)(4l_B+1)$ independent components. The number of 
measurement settings required to perform complete quantum
tomography for entangled spin systems is therefore equal to 
$(4l_A+1)(4l_B+1)$. In the experimentally
realized case of $l_A=l_B=1$ \cite{Lam01,How02,Mai01,Vaz02}, 
this would require 25 
different measurement settings with nine possible outcomes each,
for a total of 225 measurement probabilities. 

An explicit description of the density matrix in terms of 
the correlated $(n_A,n_B)$th order spin statistics can be
obtained using products of the basis operators for each 
individual system. The expansion of the density matrix
then reads
\begin{eqnarray}
\label{eq:2expand}
\hat{\rho}_{AB} &=&
\frac{1}{(2 l_A+1)(2 l_B+1)} \hat{1}\otimes \hat{1}
+ \frac{1}{2 (2 l_A+1)} \sum_{n_A} \left(\sum_i 
\langle\hat{\lambda}_{n_A,i}\otimes\hat{1}\rangle\;
\hat{\lambda}_{n_A,i}\otimes\hat{1} \right) 
\nonumber \\ &&
+ \frac{1}{2(2 l_B+1)} \sum_{n_B} \left(\sum_i 
\langle\hat{1}\otimes\hat{\lambda}_{n_B,i}\rangle\;
\hat{1}\otimes\hat{\lambda}_{n_B,i} \right)
+ \frac{1}{4} \sum_{n_A,n_B} \left(\sum_{i,j}  
\langle\hat{\lambda}_{n_A,i}\otimes\hat{\lambda}_{n_B,j}\rangle\;
\hat{\lambda}_{n_A,i}\otimes\hat{\lambda}_{n_B,j} \right).
\end{eqnarray}
The expectation values defining the density matrix can now
be expressed in terms of the joint measurement probabilities
$p_{ij}(m_A,m_B)$ by writing the $(n_A,n_B)$th order 
expectation values of the correlated spins in 
equation(\ref{eq:2stat}) as a function of the expectation 
values in equation (\ref{eq:2expand}). It is then possible
to fully characterize any NxM entanglement in terms of the
correlated spin statistics. 

The extension of this formalism to multi-partite spin 
networks is also straightforward, since the density 
matrix can be expanded into products of the local 
basis operators for any number of systems.
The expectation values of these products can then 
be determined from the correlated measurement statistics
of spin measurements performed simultaneously on all
systems.

\section{Conclusions}
In conclusion, we have shown how the density matrix of
spin-$l$ systems can be reconstructed from the measurement
statistics of projective spin measurements along a set
of at least $4l+1$ different spin directions. The 
components of the density matrix can then be identified 
with different contributions to the statistics
of the three dimensional spin vector.
It is therefore possible to interpret the discrete 
quantum statistics of arbitrarily large spin systems within the 
same three dimensional space defined by the Bloch vector 
of a two level system. 

The explicit procedure for the quantum tomography of 
spin-1 systems provides an example of the general
method that can be applied directly to experimentally
generated two photon polarization states such as the
ones reported in \cite{Tse00,Bur02,Lam01,How02}. 
It may thus serve as the foundation of 
a more detailed characterization of decoherence and
noise effects in these newly available entanglement 
sources.

Besides its practical usefulness for the experimental
characterization of general spin-$l$ systems, the expansion
of the density matrix into elements of the spin statistics
also provides a more intuitive understanding of quantum 
statistics in large systems. The analysis presented above
may therefore also help to clarify the conditions for the 
emergence of quantum effects in physical systems of 
arbitrary size.

\section*{Acknowledgements}
H.F.H. would like to thank A.G. White and K. Tsujino for
some very motivating discussions.

%As the analysis of decoherence illustrates, the higher
%order expectation values represent highly resolved
%details of the spin statistics, whereas the lower order
%expectation values provide quantitative estimates of
%the average spin and its fluctuations. 

%The physical properties associated with macroscopic
%superpositions are in fact quite microscopic.


\begin{thebibliography}{xyz00}
\bibitem{foot1}
It may be worth noting that, in terms of group theory,
the $2n+1$ expectation values of $n$th order form 
an irreducible representation of the spherical
symmetry defined by the arbitrary direction of the 
spin component $\hat{L}_i$ in equation (\ref{eq:stats}).
\bibitem{foot2}
In fact, the truncation of the measurement statistics at 
$\langle (\hat{L}_i)^{2l} \rangle$ itself is a consequence of 
the fact that the discreteness of the quantized eigenvalues 
makes a higher resolution unnecessary.
\end{thebibliography}
\end{document}